%% file: main.tex
\documentclass{article}
\usepackage{graphicx} 
\DeclareUnicodeCharacter{221A}{\'{e}} 
\DeclareUnicodeCharacter{0308}{\"{}} 

\usepackage[utf8]{inputenc}
\usepackage[T1]{fontenc}
\usepackage{amsmath}
\usepackage{amsfonts}
\usepackage{amssymb}
\usepackage[version=4]{mhchem}
\usepackage{stmaryrd}
\usepackage{xcolor}
\usepackage{cancel}
\usepackage{graphicx}
\usepackage{svg}
\usepackage{soul}

\usepackage[utf8]{inputenc}
\usepackage[T1]{fontenc}
\usepackage{graphicx}
\usepackage{subcaption}
\usepackage{geometry}
\usepackage[export]{adjustbox}
\graphicspath{ {./images/} }
\usepackage{amsmath}
\usepackage{amsfonts}
\usepackage{amssymb}
\usepackage[version=4]{mhchem}
\usepackage{stmaryrd}
\usepackage{bbold}
\usepackage{hyperref}
\hypersetup{colorlinks=true, linkcolor=blue, filecolor=magenta, urlcolor=cyan,}
\usepackage{xcolor}
\usepackage{cancel}
\usepackage{authblk}
\usepackage{pgfkeys}

\usepackage{todonotes}
\usepackage[normalem]{ulem}

\input{sample-values}

\title{Two Strategies to Measure Spin-Orbit-Torque Efficiency Acting on the Insulating Magnet \LAFO{71}{Comp}}

\author[1]{Orion Smedley}
\author[1]{Thow Min Jerald Cham}

\author[3,5]{Daisy O'Mahoney}
\author[2,3]{Sanyum Channa}
\author[3,4]{Xin Yu Zheng}
\author[2,3]{Anna Janni}
\author[3,4]{Lauren J. Riddiford}

\author[1]{Yuhan Liang}
\author[1]{Bozo Vareskic}

\author[7]{Zbigniew Galazka}
\author[6]{Yunqiu Kelly Luo}
\author[3,5]{Yuri Suzuki}
\author[1,8]{Daniel C. Ralph}

\affil[1]{Department of Physics, Cornell University, Ithaca, NY, 14853, USA}

\affil[2]{Department of Physics, Stanford University, Stanford, CA, 94305, USA}

\affil[3]{Geballe Laboratory for Advanced Materials, Stanford University, Stanford, CA, 94305, USA}

\affil[4]{Department of Applied Physics, Stanford University, Stanford, CA, 94305, USA}

\affil[5]{Department of Materials Science and Engineering, Stanford University, Stanford, CA, 94305, USA}

\affil[6]{Department of Physics and Astronomy, University of Southern California, Los Angeles, CA, 90089, USA}

\affil[7]{Leibniz-Institut f\"ur Kristallz\"uchtung,Max-Born-Str. 2, 12489 Berlin, Germany}

\affil[8]{Kavli Institute at Cornell, Cornell University, Ithaca, NY, 14853, USA}

\date{\today}

\begin{document}

\maketitle

\noindent \textbf{\Large{Abstract}} \\

\noindent For insulating magnets, widely-used electrically-based techniques to measure spin-orbit-torque efficiency can suffer from poor signal-to-noise ratios and unwanted artifacts, while optical measurements based on conventional magneto-optical Kerr effect (MOKE) imaging often lack sufficient sensitivity. Here we analyze two strategies for measuring the spin-torque efficiency acting on an insulating magnet, focusing on the torque from Pt acting on thin films of the low-damping magnet lithium aluminum ferrite (LAFO). First, lateral $+$ longitudinal spin-torque ferromagnetic resonance (ST-FMR) allows separation of the spin-torque signal from artifacts due to spin-pumping and resonant heating. With this technique, we measure a spin-torque efficiency $\xi^\text{ST-FMR}_\text{DL}\sim\LAFO{71}{XiST}$, comparable to but a bit less than for Pt acting on metallic ferromagnets. Second, we attempted optical measurements of spin-orbit-torque-induced magnetic deflection using Sagnac interferometry. We find that Sagnac interferometry provides sufficient sensitivity to measure signals from \LAFO{71}{ThkNom} LAFO thin films, but the inferred torque efficiency is much weaker than the result from lateral $+$ longitudinal ST-FMR. We speculate that there may be another contribution to the Sagnac signal other than current-induced magnetic deflection, perhaps an out-of-plane component of itinerant electron spins. This becomes relevant for insulating magnets in which the magneto-optic coupling coefficient is very weak.

\section*{Introduction}

Insulating magnetic materials provide both opportunities and challenges for spintronics experiments and potential technologies \cite{Wu_2013,brataas_spin_2020}.  Insulating magnets can possess Gilbert damping constants much lower than metallic magnets, which in principle permits more efficient magnetic manipulation by damping-like spin-orbit torque.  Low damping also allows long-distance transmission of magnons \cite{cornelissen_long-distance_2015,brataas_spin_2020}.  However, whereas metallic magnets can enable large magnetoresistance signals that allow easy read-out of the magnetic orientation, the magnetoresistance of heterostructures containing insulating magnets is generally orders of magnitude weaker. This complicates the use of magnetoresistive read-out to measure of the efficiencies of spin-orbit torque from a heavy metal acting on insulating magnets, because unintended artifacts can become comparable to the intended electrical signals generated by the spin-orbit torque. For example, No\"el et al.\ \cite{noel_nonlinear_2025,noel_estimation_2025} have shown that current-induced magnon creation and annihilation can interfere with the commonly-used spin-orbit-torque measurement technique based on second-harmonic Hall detection \cite{pi_tilting_2010,Garello_2013,HarmonicHallHayashi_2014,SOT_thermoElectric_avci_2014},
leading to as much as an order-of-magnitude error in the conventional analysis for insulating magnets.
Optical measurements of current-induced magnetic deflection based on the magneto-optical Kerr effect (MOKE) are an alternative approach to measure spin-orbit torques that is effective for metallic ferromagnets \cite{noel_estimation_2025,Fan_2014,Montazeri_2015,Fan_2016}, however conventional MOKE imaging generally lacks the sensitivity necessary to measure the spin-orbit torque efficiency for thin films of roughly 10 nm for all but a few magnetic insulators \cite{noel_estimation_2025} (e.g., in ref.\ \cite{Montazeri_2015} for Y$_3$Fe$_5$O$_{12}$ a 50 nm thick layer was required).

Here, we explore two strategies to measure the damping-like spin-orbit torque acting on \LAFO{71}{ThkNom} thin films of the low-damping magnetic insulator \LAFO{71}{Comp}: (1) lateral $+$ longitudinal spin-torque ferromagnetic resonance \cite{SabaJosephHallSTFMR} (ST-FMR), which allows the spin-torque signal to be separated from artifacts due to spin pumping and resonant heating that can interfere with conventional ST-FMR measurements and (2) Sagnac optical interferometry \cite{XiaSagnac2006,OurSagnac}, an alternative to conventional MOKE imaging with improved sensitivity.
We choose to study films from the family Li$_{0.5}$Al$_x$Fe$_{2.5-x}$O$_4$ (LAFO) because they exhibit Gilbert damping values as low as 1.5 $\times$ 10$^{-4}$ for 15 nm samples \cite{zheng_ultra-thin_2023,OMahoney2026}, comparable to similar-thickness thin films of iron garnets, the most-common family of low-damping magnetic insulators (e.g., Y$_3$Fe$_5$O$_{12}$ (YIG) and Tm$_3$Fe$_5$O$_{12}$ (TmIG) \cite{ding_nanometer-thick_2020,soumah_ultra-low_2018}).   A previous second-harmonic Hall measurement for LAFO/Pt reported a very large value for the damping-like spin-orbit torque efficiency per unit charge density $\xi^\text{2HH}_\text{DL}$ = 0.57 \cite{zheng_ultra-thin_2023}, but did not include correction for magnon creation and annihilation \cite{noel_estimation_2025}.  Here, for \LAFO{71}{Stack} samples we find using lateral $+$ longitudinal ST-FMR an efficiency $\xi^\text{ST-FMR}_\text{DL} = \LAFO{71}{XiST}$ comparable to Pt acting on metallic magnets, while Sagnac interferometry implies a value much lower. We speculate why optical measurements of spin-orbit-torque-induced magnetic deflection may be inaccurate for insulating magnets with very weak magneto-optical coupling coefficients.

\section*{LAFO thin film growth and characterization}

\begin{figure}
    \centering
    \includegraphics[width=1\linewidth]{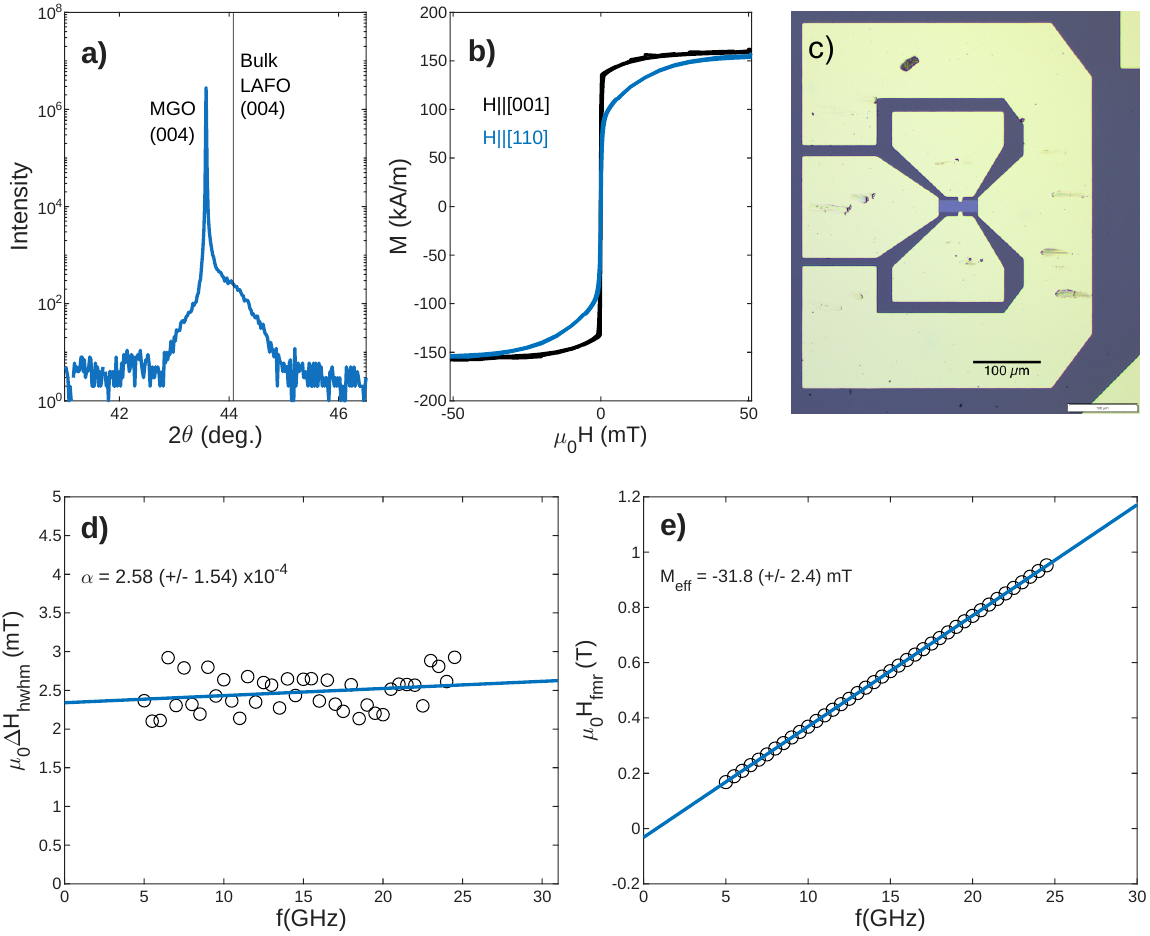}
    \caption[Structural and magnetic characterization.]{Structural and magnetic characterization of the LAFO thin films. a) X-ray diffraction of 11.3 nm LAFO on (001) oriented MGO around the (004) film peak. b) Static magnetic measurements at room temperature via SQUID magnetometry of magnetization (M) versus applied magnetic field (H) on a 11.3 nm thick LAFO sample along the in-plane [110] and out-of-plane [001] directions. c) Geometry of the contact pads used for the lateral + longitudinal ST-FMR.  Scale bar is 100 $\mu$m.  d) FMR frequency dependence of the full-width-half-maximum linewidth ($\mu_0H_\text{FWHM}$).  e) FMR resonance field ($\mu_0H_\text{FMR}$) as a function of frequency.  In panels d) and e) the magnetic field is oriented along the [001] direction.   }
    \label{fig:Characterization}
\end{figure}
Epitaxial LAFO/Pt bilayers were characterized and patterned into Hall bars. Details of the synthesis on single crystal MgGa$_2$O$_4$ (MGO) substrates and structural charqacterization can be found in Supplementary Section A and ref.~\cite{OMahoney2026}. Static magnetic properties measured by SQUID magnetometry indicate a weakly out-of-plane magnetic anisotropy (Fig.~1b). For 11.3 nm unpatterned test samples, broadband ferromagnetic resonance measurements (FMR) yield a magnetic anisotropy $\mu_0 M_\text{eff}=-32$ mT (Fig.~1e), while for the 14.4 nm lithographically-defined devices upon which the torque measurements were performed $\mu_0 M_\text{eff}=-0.12(2)$ T, based on the magnetic field required to saturate the magnetization in plane (see Supplementary Section \ref{sec: MeffCal}). This value is small enough that a small applied magnetic field can saturate the magnetization in-plane for the ST-FMR and Sagnac MOKE measurements. On representative comparison films, SQUID measurements provide a determination of the saturation magnetization $M_s = \LAFO{71}{Ms}$, and FMR measurements  give a Gilbert damping parameter $\alpha = 2.58 \times 10^{-4}$ (for bare LAFO without an adjacent Pt layer). After deposition of Pt, the damping parameter determined by ST-FMR measurements on fabricated devices is $\alpha = \LAFO{71}{AlphaST}$.

We patterned a \LAFO{71}{Stack} sample into ST-FMR Hall bars with a channel width of \LAFO{71}{ChanW} and length of \LAFO{71}{DevL}. Platinum contacts were deposited in the ST-FMR Hall geometry introduced in ref.\ \cite{SabaJosephHallSTFMR} (Fig.\ \ref{fig:Characterization}c).

\section*{Lateral $+$ longitudinal spin-torque ferromagnetic resonance}
For conventional longitudinal ST-FMR \cite{Liu2011}, spin pumping and resonant heating \cite{SpinPump1,SpinPump2,SpinPump3,Heating1} can contribute voltage signals with the same dependence on the angle of an in-plane magnetic field as the signals due to spin-orbit torques, thereby interfering with accurate measurements of spin-orbit-torque efficiency.  
The challenge of separating out these artifacts is potentially even more problematic for samples containing insulating magnets, compared to metallic magnets, because the magnetoresistance signals used to detect the spin-orbit torque can be orders of magnitude smaller for insulating magnets.  Nevertheless, here we show that the method of lateral $+$ longitudinal ST-FMR \cite{SabaJosephHallSTFMR,ChamHallSTFMR} can still be used to separate out artifact voltages and make quantitative measurements of spin-orbit-torque efficiency for Pt/LAFO samples. 

\begin{figure}
    \centering
    \includegraphics[width=0.75\linewidth]{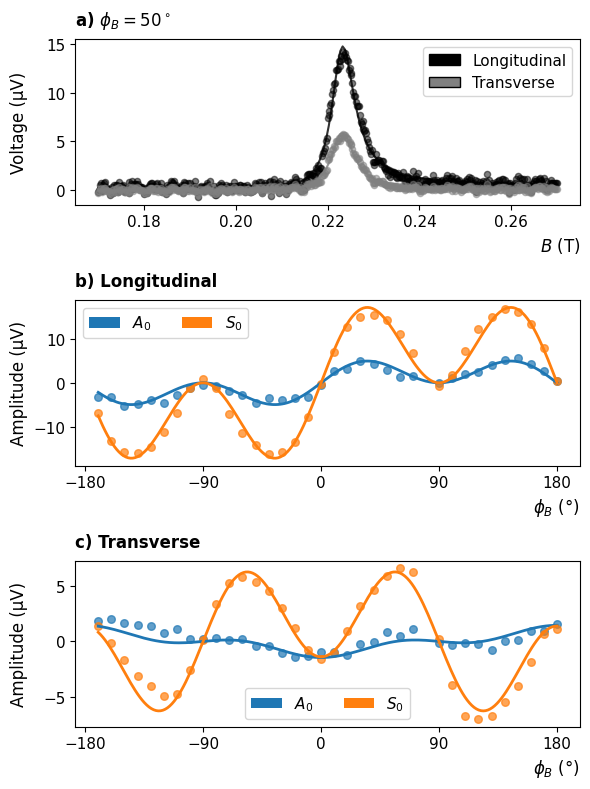}
    \caption{
    Measurement of the damping-like spin--orbit torque using spin-torque ferromagnetic resonance (ST-FMR) for a \LAFO{71}{Stack} sample at room temperature.
    (a) Representative field sweeps at $\phi=50^{\circ}$ with Lorentzian fits to the longitudinal and transverse (Hall) DC mixing voltages.
    The extracted symmetric (orange) and antisymmetric (blue) components are plotted as a function of $\phi$ for the (b) longitudinal and (c) transverse signals.
    The angular dependence (Eq.~(\ref{eq:stfmr_longV})) is fitted to obtain the coefficients listed in Table~\ref{table:coefficients}, which are used to quantify the spin--orbit torque amplitudes.
    \label{fig:71STFMR}
    }
\end{figure}

We performed the ST-FMR measurements on a \LAFO{71}{Stack} chip (sample length $L=\LAFO{71}{DevL}$, channel width $W_\text{channel}=\LAFO{71}{ChanW}$, distance between the ends of the Hall probes $W=\LAFO{71}{HallW}$). A \LAFO{71}{Frf} microwave signal was applied along the length of the Hall bar with an initial incident power of \LAFO{71}{Prf_src}. The microwave current within the sample was calibrated by comparing the microwave-induced heating to heating by a direct current (see Supplementary Section B). For a \LAFO{71}{Prf_src} incident signal, the calibration indicates that the peak microwave current flowing within the sample is $I_\text{rf}=\LAFO{71}{Irf_pk_samp}$. 
This RF current will excite magnetic precession via both its Oersted magnetic field and spin-orbit torque, leading to DC voltages arising from the combination of rectified resistance oscillations, spin-pumping voltages, and possibly resonant heating.  Both the longitudinal DC voltage and the Hall DC voltage (Fig.\ \ref{fig:71STFMR}) have the form of a sum of a symmetric Lorentzian ($\propto S(B)=\Delta^2/[(B-B_0)^2+\Delta^2]$) and an antisymmetric Lorentzian ($\propto A(B)=(B-B_0)\Delta/[(B-B_0)^2+\Delta^2]$), where $\Delta$ is the line width and $B_0$ is the resonant field. Specifically, for a high-symmetry spin-source layer like Pt, according to reference \cite{SabaJosephHallSTFMR}:
\begin{equation}
\begin{aligned}
V_{XX}=\sin(2\phi)\cos(\phi)\big[S^\text{AMR/art}_{XX}S(B)+A^\text{AMR}_{XX}A(B)\big]
\label{eq:stfmr_longV}
\end{aligned}
\end{equation}
\begin{equation}
\begin{aligned}
V_{XY}=\cos(2\phi)\cos(\phi)\big[S^\text{PHE/art}_{XY}S(B)+A^\text{PHE}_{XY}A(B)\big] + \cos(\phi)\big[S^\text{AHE/art}_{XY}S(B)+A^\text{AHE}_{XY}A(B)\big]
\label{eq:stfmr_HallV}
\end{aligned}
\end{equation}
with the coefficients
\begin{equation}
\begin{aligned}
S_{X X}^{\mathrm{AMR}/\mathrm{art}} & =\frac{I_{\mathrm{rf}}}{2 \alpha \omega^{+}} R_{\mathrm{AMR}} \tau_{\mathrm{DL}}^0-\frac{L}{2} E_{\mathrm{art}}^0 \\
A_{X X}^{\mathrm{AMR}} & =\frac{I_{\mathrm{rf}}}{2 \alpha \omega^{+}} R_{\mathrm{AMR}} \frac{\omega_2}{\omega} \tau_z^0, \\
S_{X Y}^{\mathrm{PHE}/\mathrm{art}} & =-\frac{I_{\mathrm{rf}}}{2 \alpha \omega^{+}} R_{\mathrm{PHE}} \tau_{\mathrm{DL}}^0+\frac{W}{2} E_{\mathrm{art}}^0, \\
A_{X Y}^{\mathrm{PHE}} & =-\frac{I_{\mathrm{rf}}}{2 \alpha \omega^{+}} R_{\mathrm{PHE}} \frac{\omega_2}{\omega} \tau_z^0, \\
S_{X Y}^{\mathrm{AHE}/\mathrm{art}} & =\frac{I_{\mathrm{rf}}}{2 \alpha \omega^{+}} R_{\mathrm{AHE}} \tau_z^0+\frac{W}{2} E_{\mathrm{art}}^0, \\
A_{X Y}^{\mathrm{AHE}} & =-\frac{I_{\mathrm{rf}}}{2 \alpha \omega^{+}} R_{\mathrm{AHE}} \frac{\omega_1}{\omega} \tau_{\mathrm{DL}}^0 .
\label{eq: stfmr angle coef}
\end{aligned}
\end{equation}
We measured the dependence of $V_{XX}$ and $V_{XY}$ as a function of changing the angle $\phi$ of the in-plane magnetic field (Fig.\ \ref{fig:71STFMR}b,c) using two lock-in amplifiers with sinusoidal modulation to 100\% depth of the applied RF amplitude, and fit to the $\phi$ dependence to determine the coefficients $S_{X X}^{\mathrm{AMR}/\mathrm{art}}$, $A_{X X}^{\mathrm{AMR}}$, $S_{X Y}^{\mathrm{PHE}/\mathrm{art}}$, $A_{X Y}^{\mathrm{PHE}}$, $S_{X Y}^{\mathrm{AHE}/\mathrm{art}}$, and $A_{X Y}^{\mathrm{AHE}}$ (Table S1).

In this treatment, we use the conventions for the longitudinal resistance  $R_{XX}=R_0+R_{\mathrm{AMR}} m_X^2$ and the Hall resistance $R_{XY}=R_\text{PHE}m_Xm_Y+R_\text{AHE}m_Z$, where
 $m_X$ is the component of the magnetization unit vector pointing along the current, $m_Y$ is the in-plane component perpendicular to the current, and  $m_Z$ the out-of-plane component.  For the sign of the Hall voltages, we pick the convention so that $R_\text{PHE}$ has the same sign as $R_\text{AMR}$, flipped relative to
 reference  \cite{SabaJosephHallSTFMR}, which is the reason that some of the signs in Eq.\ (\ref{eq: stfmr angle coef}) are reversed compared to \cite{SabaJosephHallSTFMR}.
 $\tau^0_\text{DL}$ is the maximum amplitude of the in-plane damping-like torque (in frequency units) that we wish to determine, $\tau^0_z$ is the maximum amplitude of the out-of-plane field-like torque (including both Oersted and spin-orbit contributions), and $E^0_\text{art}$ is the artifact electric field perpendicular to the magnetization resulting from spin pumping and/or resonant heating.  $\alpha$ is the Gilbert damping coefficient,   $\omega_{1} = \gamma B_{0}$,
$\omega_{2} = \gamma\!\left(B_{0} + \mu_{0} M_{\mathrm{eff}}\right)$,
$\omega^{+} = \omega_{1} + \omega_{2}$, and the resonance frequency $\omega=\sqrt{\omega_1\omega_2}$.  In analyzing the data, we use $\mu_0M_\text{eff}=\LAFO{71}{Meff}$ determined by saturating the magnet in plane in Supplementary Section \ref{sec: MeffCal}, and we use  $\alpha=\LAFO{71}{AlphaST}$, determined from the measured resonance linewidths according to $\alpha=\gamma\Delta/\omega$. The LAFO possesses a weak 4-fold in-plane magnetic anisotropy, with the resonant field $B_0$ at \LAFO{71}{Frf} varying over \LAFO{71}{ResBRange}; our values for $B_0$ and $\alpha$ represent averages over $\phi$. For the sample under test, $R_\text{AMR}=\LAFO{71}{RAMR}$, $R_\text{PHE}=\LAFO{71}{RPHE}$, $R_\text{AHE}=\LAFO{71}{RAHE}$, and $R_0=\LAFO{71}{R0}$, corresponding to an average Pt resistivity of \LAFO{71}{RhoPt} for the \LAFO{71}{PtThk} Pt layer.

\begin{table}
\begin{center}
    \begin{tabular}{|c | c|}
        \hline
        ST-FMR component & value  \\
        \hline
        $S^\text{AMR/art}_{XX}$ & \LAFO{71}{SamrFit} \\ [1ex]
        $A^\text{AMR}_{XX}$ & \LAFO{71}{AamrFit} \\ [1ex]
        $S^\text{PHE/art}_{XY}$ & \LAFO{71}{SpheFit} \\ [1ex]
        $A^\text{PHE}_{XY}$ & \LAFO{71}{ApheFit} \\ [1ex]
        $S^\text{AHE/art}_{XY}$ & \LAFO{71}{SaheFit} \\ [1ex]
        $A^\text{AHE}_{XY}$ & \LAFO{71}{AaheFit} \\ [1ex]
        \hline
    \end{tabular}
    \caption{Amplitudes of the components contributing to the ST-FMR signals, obtained from fits to the data in Fig.~\ref{fig:71STFMR}. Uncertainties represent the fit standard errors. }
\label{table:coefficients}
\end{center}
\end{table}

The value of $A^\text{AHE}_{XY}$ is much smaller than the other terms, reflecting that the anomalous Hall resistance $R_\text{AHE}$ is more than a factor of 30 smaller than $R_\text{PHE}$. Approximating $R_\text{AHE} \approx 0$, the equation for $S^\text{AHE/art}_{XY}$ therefore allows a determination of the artifact contribution  $(W/2)E^0_\text{art}=\LAFO{71}{SaheFit}$ , and then from the $S^\text{PHE/art}_{XY}$ equation we reach a simple expression by which we can solve for the damping-like torque:
\begin{equation}
\frac{I_{\mathrm{rf}}}{2 \alpha \omega^{+}} R_{\mathrm{PHE}} \tau_{\mathrm{DL}}^0 = -S_{X Y}^{\mathrm{PHE/art}} +S^\text{AHE/art}_{XY} = \LAFO{71}{VPheDL}.
\label{almosttorque}
\end{equation}
or alternatively,
\begin{equation}
\frac{I_{\mathrm{rf}}}{2 \alpha \omega^{+}} R_{\mathrm{AMR}} \tau_{\mathrm{DL}}^0 = S_{XX}^{\mathrm{AMR/art}} +\frac{L}{W} S^\text{AHE/art}_{XY} = \LAFO{71}{VAmrDL}.
\label{almosttorque2}
\end{equation}
(If $R_\text{AHE}$ had not been so small, then by the procedure in ref.\ \cite{SabaJosephHallSTFMR} the artifact contribution could be determined via a quadratic equation.  Our approximation agrees with the full solution of this quadratic equation.)

Substituting the measured values of parameters into Eq.\ \eqref{almosttorque} or \eqref{almosttorque2}, the strength of the damping-like torque (in frequency units) is $\tau^0_\text{DL}=\LAFO{71}{TauDLPHE}$ or $\tau^0_\text{DL}=\LAFO{71}{TauDLAMR}$, respectively. Finally, the damping-like torque efficiency can be determined as
\begin{equation}
    \xi_\text{DL}=\tau^0_\text{DL} \frac{eM_st_\text{mag}}{\mu_BJ_e}
\end{equation}
where $t_\text{mag}=\LAFO{71}{Thk}$ is the thickness of the LAFO film and $J_e=\LAFO{71}{JPt}$ is the applied current density within the Pt layer. We find $\xi^\text{ST-FMR(PHE)}_{\mathrm{DL}}=\LAFO{71}{XiSTPHE}$ or $\xi^\text{ST-FMR(AMR)}_{\mathrm{DL}}=\LAFO{71}{XiSTAMR}$ from the PHE- and AMR-based extractions, respectively. The reported error is the combined fit uncertainty and the uncertainty in physical parameters that enter into the calculation. For a sample with a higher iron concentration, \LAFO{404}{Comp}, we obtain a slightly smaller torque efficiency, \LAFO{404}{XiST}, from both the PHE- and AMR-based measurements.

For comparison, an analysis of the longitudinal ST-FMR signal without accounting for the spin-pumping voltage would give  $\xi^\text{incorrect}_{\mathrm{DL}}=\LAFO{71}{XiSTWrong}$, so the spin-pumping signal represents more than a \LAFO{71}{ArtifactPct} correction. ST-FMR also allows a determination of the field-like component of the current-induced torque, which is dominated by the Oersted field as discussed in Supplementary Section D.

The value of $\xi^\text{ST-FMR}_{\mathrm{DL}}$ we determine from the lateral $+$ longitudinal ST-FMR is much less than the previous value from a second harmonic Hall measurement, $\xi^\text{2HH}_\text{DL}$ = 0.57 \cite{zheng_ultra-thin_2023}.  Based on the analysis in refs.~\cite{noel_nonlinear_2025,noel_estimation_2025}, a more accurate estimate of the spin-orbit torque efficiency from second harmonic Hall measurements requires taking into account magnon creation and annihilation.  One might also ask whether magnon creation and annihilation might compromise the ST-FMR measurement.  We assert that it should not have a significant impact, because in our ST-FMR measurements there is no large applied dc current to transduce a resistance change arising from magnon creation or annihilation to a voltage signal.  We discuss this in more detail in the Supplementary Section G.

Our results for LAFO/Pt can also be compared to previous measurements of spin-orbit torques from Pt acting on metallic ferromagnets and other insulating ferrimagnets. Because the electrical resistivities ($\rho$) of the Pt in different experiments depend on the film thickness and deposition conditions, it is useful to consider the spin torque conductivity, $\sigma_\text{DL}=\xi_\text{DL}/\rho$ (i.e., the torque efficiency per unit applied electric field) as well as the spin torque effiency per unit applied current density, $\xi_\text{DL}$; based on our measurements on \LAFO{71}{Stack} with  $\rho=\LAFO{71}{RhoPt}$ we have $\sigma^\text{ST-FMR}_\text{DL}=\LAFO{71}{SigmaSH}$.
The spin-orbit torque efficiency from Pt acting on metallic magnets such as permalloy, Fe$_{60}$Co$_{20}$B$_{20}$, or Co is
$\xi^\text{metals}_\text{DL}$ = 0.09-0.12  \cite{SabaJosephHallSTFMR,OurSagnac}, corresponding to $\sigma^\text{metals}_\text{DL}\sim 5 \times 10^5$  $(\hbar/2e)$ $\Omega^{-1}$m$^{-1}$ \cite{ZhuPtReview}. Our LAFO results are likely reduced in part compared to the measurements on metallic ferromagnets because our small Pt thickness, \LAFO{71}{PtThk}, is likely not much larger than the spin diffusion length, but still the reduced values we find suggest that the spin torque conductivity for our LAFO/Pt interfaces is less than for a typical Pt/(metallic ferromagnet) interface. Our measurements are slightly above the upper end of the range for the spin-torque efficiencies for Pt acting on insulating low-damping iron garnets as measured by other techniques, $\xi_\text{DL}^\text{garnets} = 0.01 - 0.06$ \cite{avci_current-induced_2017,li_deficiency_2017,avci_APL2017,guo_spin-orbit_2019,LiNanomaterials2023}.

\section*{Sagnac interferometry measurements}
Our procedure for using Sagnac interferometry to measure spin-orbit-torque efficiency is to apply a direct charge current to a magnet/heavy-metal bilayer to reorient the magnetization, and then to use Sagnac interferometry to quantify the out-of-plane magnetic deflection \cite{OurSagnac}. In a Sagnac interferometer, two circularly-polarized light beams related by time reversal are incident on the sample and the phase difference between them is measured, so that only effects which break time reversal symmetry contribute to the signal.  This selectivity  rejects thermal artifacts and boosts Sagnac's sensitivity to be 50-100x better than traditional unmodulated MOKE measurements \cite{OurSagnac}.

We used the same apparatus and measurement protocol our research group described previously for a study of Pt/Co samples \cite{OurSagnac}. We adjusted the super-luminescent diode light source so that the average light intensity reaching the photoreceiver was $\approx\LAFO{71}{Poptic}$. This ensured that the detector signal was not so large that it risked being clipped at peak modulation amplitudes. We aligned our laser spot to be in the center of the Hall bar in between the Hall leads. All quantitative measurements were done as follows. We demodulated with a Zurich HF2LI lock-in amplifier set to a 1st order filter. After each change in the measurement conditions, we allowed a settling time equal to \LAFO{71}{SettleN} lock-in time constants. At each field point, we acquired successive measurements over a fixed averaging interval before stepping to the next field value. The time constant, averaging interval per field point, and the number of repeated field sweeps are given in the corresponding figure captions.
We operated the electro-optic modulator at the frequency $\omega=\LAFO{71}{FEOM}$ tuned to the optical path length. To drive the spin-orbit torque, the lockin amplifier applied an alternating current to the longitudinal leads with amplitude \LAFO{71}{SagIpk} (\LAFO{71}{SagIrms})c at $\omega_e=\LAFO{71}{FDrive}$, a frequency chosen to have few common divisors with respect to $\omega$, so that its harmonics are far from the electro-optic modulation frequency.

Compared with Ref.~\cite{OurSagnac}, our data acquisition differed in two respects. First, we recorded the current-modulated signal $(\Delta\theta_k)$ at both sidebands and average their difference, measuring $(I^{\omega-\omega_e}-I^{\omega+\omega_e})/2$, where $I^{\omega+\omega_e}$ and $I^{\omega-\omega_e}$ are the intensities detected at the upper and lower sidebands, respectively. This reduced the measurement noise by a factor of $\sqrt{2}$. Second, we used a more precise technique to set the phases on the lockin amplifier. To do this we phased the second harmonic so that the signal lies entirely in the quadrature component; then we set the first-harmonic phase to be half of the second-harmonic phase, assuming the phase difference arises from a constant signal delay. For the sidebands, we determined the phase in post-processing from a linear fit to the in-phase and quadrature components, because our Zurich Instruments HF2LI was not equipped with the optional modulation module. This phase determination has a 180° ambiguity; consequently, we report only the magnitude of the Sagnac-inferred torque efficiency.

\begin{figure}
    \centering
    \includegraphics[width=1\linewidth]{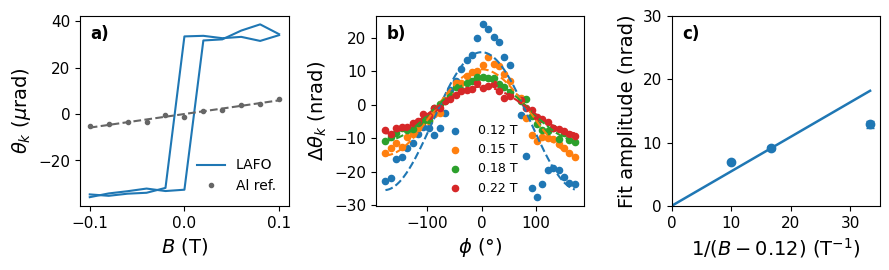}

    \caption{Attempt to measure damping-like spin-orbit torque using Sagnac interferometry for a \LAFO{71}{Stack} sample at room temperature. (a) Calibration of the magneto-optic coupling constant $\kappa$. The Kerr signal is measured versus out-of-plane field for the sample and an aluminum mirror background reference.
    (b) Measurements of current-induced Kerr rotation as a function of the applied in-plane magnetic-field angle. 
    (c) Amplitude of the oscillations in $\Delta \theta_k$ versus $\phi$ for different magnitudes of magnetic field $B$, with a linear fit to $1/(B+\mu_0 M_\text{eff})$ using the previously-determined value $\mu_0 M_\text{eff}=\LAFO{71}{Meff}$. Here, we use a time constant $\tau=\LAFO{71}{TauSag}$,  acquisition interval \LAFO{71}{Tacquire}, and an average over \LAFO{71}{SweepN} sweeps.
    }
    \label{fig: 404Sagnac}
\end{figure}

The first step in using a Sagnac interferometer to measure current-induced magnetic deflection is to calibrate the magneto-optic coupling coefficient $\kappa$ for a given magnetic sample, defined such that
\begin{equation}
\theta_k=\kappa m_Z
\label{eq: kappadef}
\end{equation}
where $\theta_k$ is the Kerr angle measured by the Sagnac interferometer and $m_Z$ is the out-of-plane component of the magnetization unit vector.
Figure~\ref{fig: 404Sagnac}(a) shows the Sagnac signal as a function of swept out-of-plane magnetic field for the same \LAFO{71}{Stack} sample discussed in the lateral $+$ longitudinal ST-FMR section, along with a calibration signal from an aluminum mirror with no magnet present. The Sagnac signal from the LAFO is weak enough that we take the small linear background into account. (This background is likely due to a Verdet contribution from the objective.) Based on the jump in the Sagnac signal upon reversal of the magnet we determine $\kappa=\LAFO{71}{Kappa}$.
This value is orders of magnitude weaker than the magneto-optic coupling for metallic magnets. For example, for the nm-scale Co films studied in ref.\ \cite{OurSagnac} $\kappa = 8000$ $\mu$rad, more than 2 orders of magnitude larger. Nevertheless, the high sensitivity of the Sagnac interferometer still allows for measurements of the current-induced MOKE signals in our \LAFO{71}{ThkNom} LAFO film.

The second step for extracting the torques is to measure the current-induced deflections in the magnetization by applying an AC current through the platinum layer and measuring the modulation of the Kerr rotation signal $\Delta\theta_k$. We attempt to measure the damping-like component of the spin-orbit torque by detecting the current-induced out-of-plane deflection of the magnetization for samples in which an applied in-plane magnetic field initializes the magnetization in-plane \cite{OurSagnac}. Figure \ref{fig: 404Sagnac}b shows the Sagnac signal for the \LAFO{71}{Stack} sample as a function of the angle $\phi$ of an in-plane magnetic field, for a selection of field strengths and an AC current amplitude $I_{rms}=\LAFO{71}{SagIrms}$ (corresponding to an RMS current density in the Pt of $J_e=\LAFO{71}{SagJ}$). The high crystal symmetry of Pt requires that the spin-orbit torque has only a conventional component ($\propto \cos \phi$, where $\phi$ is the in-plane angle of the applied magnetic field relative to the current direction). Consequently, the relation between the current-induced change in the Kerr angle and the the damping-like torque efficiency should be \cite{OurSagnac}
\begin{equation}
\Delta \theta_k=\xi^\text{Sagnac}_\text{DL}\kappa\frac{\mu_BJ_e}{\gamma eM_s t_\text{mag}}\frac{\cos(\phi)}{B+\mu_0M_{\mathrm{eff}}}
\label{eq: sagnacIP}
\end{equation}
where
\(B\) is the applied magnetic field. Figure \ref{fig: 404Sagnac}b shows good agreement with the expected $\cos \phi$ angular dependence.

A first indication that the Sagnac measurement yields unexpected results comes from an analysis of the signal amplitude as a function of magnetic field strength (Fig.~\ref{fig: 404Sagnac}(c)). The data point for the 0.15 T measurement departs from the expected $(B+\mu_0M_{\mathrm{eff}})^{-1}$ dependence, using the value $\mu_0M_{\mathrm{eff}}=\LAFO{71}{Meff}$.  (The data point for 0.12 departs even more strongly, but this field is sufficiently close to $|\mu_0 M_\text{eff}|$ that the sample may not be fully saturated in-plane.) If we fit the data points 
to Eq.~(\ref{eq: sagnacIP}), we find $|\xi^\text{Sagnac}_\text{DL}|=\LAFO{71}{XiSagFixed}$. If we attempt to allow $\mu_0M_{\mathrm{eff}}$ to vary as an adjustable parameter, we obtain $|\xi^\text{Sagnac}_\text{DL}|=\LAFO{71}{XiSagFloat}$ and a value for $\mu_0M_{\mathrm{eff}}$, $\LAFO{71}{MeffSagFit}$, that is clearly incompatible with independent characterization methods (see Supplementary Section C). For comparison, a sample with similar thickness but higher iron content, \LAFO{404}{Comp}, yielded a somewhat larger (but still small) Sagnac-inferred torque efficiency of \LAFO{404}{XiSag}.

These determinations of $\xi^\text{Sagnac}_\text{DL}$ are all much weaker than our ST-FMR result $\xi^\text{ST-FMR}_\text{DL}$, by more than an order of magnitude.  The Sagnac values are also inconsistent with experiments showing that spin-orbit-torque from Pt can drive steady-state nano-oscillations in similar Pt/LAFO samples at reasonable current densities \cite{RenAutoOscillations2025}.  We infer that the Sagnac values are incorrect, and our analysis must be neglecting important physics related to the Sagnac signal.  We speculate that the out-of-plane tilting of the LAFO magnetization is not the only source of Kerr rotation measured by the Sagnac interferometer.  One possibility is that a net out-of-plane tilting of the spins of itinerant electrons in the Pt may also contribute.  A component in the out-of-plane direction for the net spin polarization of the itinerant spins should naturally result from the combination of precession in the exchange field of the LAFO of the in-plane-oriented spin current generated by the spin Hall effect together with relaxation due to spin-orbit scattering in the Pt.   When measuring the spin-orbit torque acting on magnets with very small magneto-optical coefficients, such as LAFO, we speculate that this net canting of the itinerant electron spins might generate a Sagnac signal opposite in sign to the signal from the tilting of the LAFO magnetization.  Weighing against this suggestion is that in order to largely cancel the signal from the tilting of the LAFO magnetization the artifact signal would need to be rather large.  The Kerr rotation produced by the conventional in-plane spin accummulation associated with the spin-orbit torque in Pt films (with no adjacent magnet) has been measured in the range 7 - 50 nrad/($10^7$ A/cm$^2$) \cite{erve2014,riego2016,su2017,ervearxive,stamm2017,Marui2023,Vavassori2026}. The artifact in our measurement would need to produce a signal of order 250 nrad/($10^7$ A/cm$^2$), requiring that the MOKE signal in our devices be enhanced compared to bare Pt films by, e.g., the larger resistivity of our samples compared to the thicker films used in the earlier optical measurements and/or a proximity effect in our Pt/LAFO structures.

\section*{Conclusions}
We have investigated two strategies for measuring the efficiency of spin-orbit torque from Pt acting on the low-damping insulating magnet \LAFO{71}{Comp} (LAFO). Using Sagnac interferometry we find a physically non-sensible result, suggesting that current-induced changes in Kerr rotation can be affected by physics other than the simple tilting of the LAFO magnetization. We speculate that precession of the spins of transient electrons within the Pt due to the exchange field of the LAFO might contribute significantly to the Kerr signal, given the very weak magneto-optical coupling of the LAFO. Using the technique of lateral $+$ longitudinal ST-FMR we find for \LAFO{71}{Stack} that $\xi^\text{ST-FMR}_\text{DL}=\LAFO{71}{XiST}$, which we view as more trustworthy, since it is comparable to the values found for the spin-orbit torque acting on metallic ferromagnets. (A first-principle calculation suggests that the spin transparency for a well-ordered interface should be similar for magnetic insulators and metals \cite{jia_spin_2011}.) Our value is at the high end of the range of results determined previously using other measurement techniques for Pt acting on insulating low-damping iron garnets, $\xi_\text{DL}^\text{garnets} = 0.01 - 0.06$ \cite{avci_current-induced_2017,li_deficiency_2017,avci_APL2017,guo_spin-orbit_2019,LiNanomaterials2023}.
A torque efficiency of $\xi^\text{ST-FMR}_\text{DL}=\LAFO{71}{XiST}$ is sufficient to support LAFO as  a promising material for studies in which it is desirable to integrate thin films of a low-damping magnetic insulator with spin-orbit-torque materials to investigate the physics of magnon generating, transmission, control, and detection \cite{RenAutoOscillations2025}.  Additional improvements may also be possible from further optimization of the LAFO/Pt interface.

\section*{Acknowledgments}
We thank Xiaoxi Huang for technical assistance. This work was supported as part of the Center for Energy Efficient Magnonics, an Energy Frontier Research Center funded by the U.S. Department of Energy, Office of Science, Basic Energy Sciences under Award DE-AC02-76SF00515. O.S.\ acknowledges partial support from a New Frontier Award from the Arts \& Sciences College of Cornell University. D.O.\ and S.C., who synthesized the \LAFO{71}{Comp} thin films, were supported by the U.S.\ Department of Energy Office of Science, Office of Basic Energy Sciences, Division of Materials Sciences and Engineering under Contract No.\ DESC0008505. This work was performed in part at the Cornell NanoScale Facility, an NNCI member supported by NSF Grant NNCI-2025233 and at the shared facilities of the Cornell Center for Materials Research. X-ray diffraction and X-ray reflectivity were performed in the shared facilities of nano@stanford RRID:SCR 026695. Instrumentation support was provided by the Kavli Institute at Cornell.

\bibliographystyle{unsrtmod}
\bibliography{refs2.bib}  

\newpage
\appendix



\section{Growth and characterization of the LAFO thin films}
We synthesized \Growth{gen}{ThkRange} thick epitaxial films of \LAFO{71}{Comp} (LAFO) by pulsed laser deposition with a KrF laser operating at $\lambda = \Growth{71}{LaserLambda}$ on (001)-oriented single crystal MgGa$_2$O$_4$ (MGO)
substrates. The MGO substrates were of size \Growth{71}{SubSize} and were prepared
from high quality bulk single crystals grown by the Czochralski
method at the Leibniz-Institut f\"ur Kristallz\"uchtung, Berlin, Germany \cite{Galazka}.  A pressed \Growth{71}{TargetComp} target was used for ablation, which includes an additional \Growth{71}{LiExcess} Li enrichment
to compensate for Li loss during deposition due to Li volatility. The deposition was then performed in \Growth{71}{O2Press} O$_2$
atmosphere with a substrate temperature of \Growth{71}{Temp}, target-to-substrate distance of \Growth{71}{TargetDist}, and a laser fluence of \Growth{71}{Fluence} operating at \Growth{71}{LaserRate}. The laser spot size was \Growth{71}{SpotArea}. After deposition, the substrate was left to cool to ambient temperature in \Growth{71}{CoolPress} O$_2$. Subsequently a \LAFO{71}{PtThk} Pt layer was deposited in a separate electron beam evaporation system on top of the LAFO film. The Pt is kept thin because the Sagnac optical measurements are performed through the Pt layer.

X-ray diffraction measurements of the (004) LAFO peak on (001) oriented MGO substrate are difficult to resolve due to the LAFO peak being directly under the substrate peak. We can assume epitaxial growth of LAFO on MGO based on previous work showing a more highly strained composition of LAFO on MGO retaining high crystallinity and epitaxy up to at least 22.5 nm\cite{zheng_ultra-thin_2023}.

\section{RF current calibration for the ST-FMR measurement} \label{sec: iCal}
\begin{figure}
    \centering
    \renewcommand{\thefigure}{S1}
    \includegraphics[width=1\linewidth]{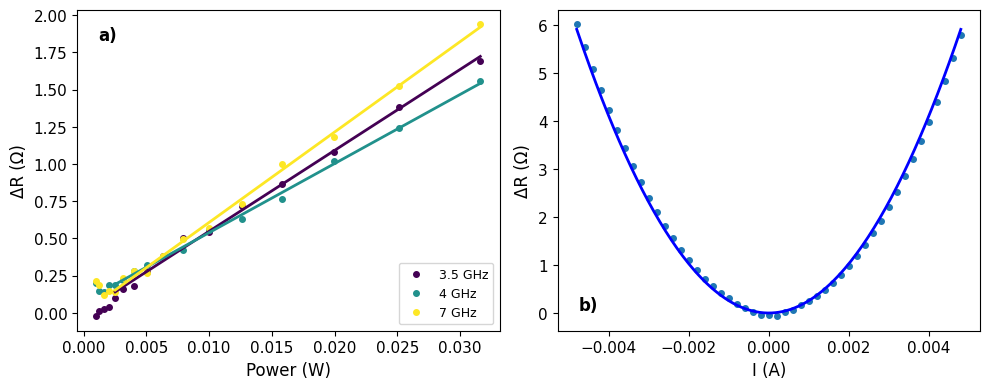}
    \caption{Calibration of RF current for ST-FMR. (a) RF current calibration, with linear fits. We repeat with different frequency of the RF current to see the deviation of the transfer factors as a function of frequency. For the ST-FMR measurements, the relevant frequency is \LAFO{71}{Frf}. (b) dc current causes thermal changes in resistance, overlayed with a parabolic fit $R=aI^2$.}
    \label{fig:iCal}
\end{figure}

To calibrate the microwave (RF) current used for the ST-FMR measurement, we compare the thermal increase in resistance induced by an RF current (Fig.\ \ref{fig:iCal}a) with the result of applying a known direct current (Fig.\ \ref{fig:iCal}b). The RF current is not modulated during this calibration.

For the RF heating we fit \(\Delta R = m P_0 \), and for the dc calibration we fit \(\Delta R = aI_\text{dc}^{2}\), where $\Delta R$ is the heating-induced resistance change, $P_0$ is the applied RF power at the RF source (linear scale), $I_\text{dc}$ is the dc current, and the other variables are fit parameters. At \LAFO{71}{Frf}, we obtain $m=\LAFO{71}{RFHeatSlope}$ and $a=\LAFO{71}{DCHeatCoeff}$.
The dc current which gives heating equivalent to an applied RF power $P_0$ is then $I_\text{dc}=\sqrt{\tfrac{m}{a}P_0}$ so that the calibration for the RF current amplitude within the sample which gives the equivalent heating is $I_\text{rf}=\sqrt{2\tfrac{m}{a}P_0}$. We can express this as a current transfer factor from the RF source to the sample for \LAFO{71}{Frf} microwaves of $I_\text{rf}/I_0=\sqrt{\tfrac{\text{m}}{a}}\sqrt{Z_{0}}=\LAFO{71}{RFTransfer}$, where $I_0$ is the amplitude of current at the source and $Z_0=\General{SrcZ}$. This transfer factor is just under what would be expected from the impedance mismatch alone, $\frac{2Z_0}{Z_0+R_L}=\LAFO{71}{MismatchTransfer}$, where the sample resistance is \(R_L=\LAFO{71}{R0}\). This seems reasonable given attenuation in the coaxial cable and loss in the connectors.
The final calibrated value of $I_\text{rf}$ for the \LAFO{71}{Frf} ST-FMR measurement, at $P_0=\LAFO{71}{Prf_src}$, is $I_\text{rf}= \LAFO{71}{RFTransfer} \sqrt{2P_0/(\General{SrcZ})} = \LAFO{71}{Irf_pk_samp}$.

\section{Determination of the Magnetic Anisotropy Strength} \label{sec: MeffCal}
\begin{figure}
    \centering
    \includegraphics[width=0.75\linewidth]{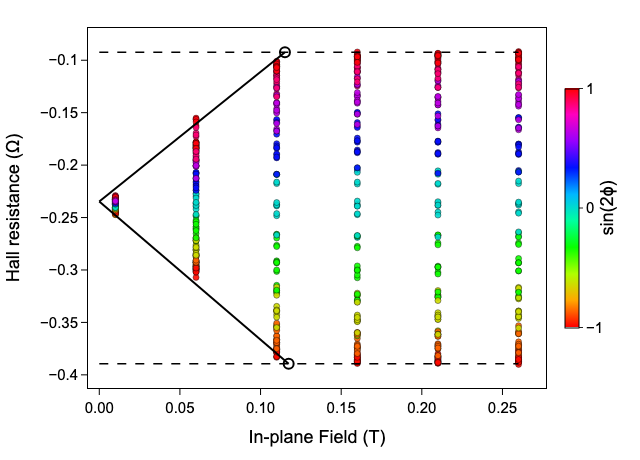}
    \caption{Planar Hall measurement of $\mu_0M_{eff}$. An applied in-plane magnetic field rotates the magnetization toward the film plane. Because the planar Hall resistance is much larger than the anomalous Hall resistance ($R_{\mathrm{PHE}} \gg R_{\mathrm{AHE}}$), the measured Hall resistance depends strongly on the in-plane field direction. We take the saturation field as our measurement of  $\mu_0M_{eff}$}
    \label{fig: Meff}
\end{figure}

The magnetic anisotropy, i.e. the effective magnetization, is $\mu_0M_{eff}=\LAFO{71}{Meff}$ as determined by the hard-axis (in-plane) field that saturates the planar Hall signal (fig. \ref{fig: Meff}). 


\section{Field-like torque determined from ST-FMR}

For the \LAFO{71}{Stack} ST-FMR sample discussed in the main text, the field-like spin-orbit torque in combination with the Oersted field can be measured from $A_\text{XX}^\text{AMR}$ in Eq.~\eqref{eq: stfmr angle coef}, without requiring any artifact correction, because the antisymetric ST-FMR components are free from spin-pumping and other artifacts that do not depend on the phase of the magnetization precession. Given the value of $A_\text{XX}^\text{AMR}$ in Table \ref{table:coefficients}, we have
$$ \tau_z^0 =A_{X X}^{\mathrm{AMR}}/ \bigg(\frac{I_{\mathrm{rf}}}{2 \alpha \omega^{+}} R_{\mathrm{AMR}} \frac{\omega_2}{\omega} \bigg)=\LAFO{71}{TauFLAMR}.$$
This can be compared to the out-of-plane torque (in frequency units) due to the in-plane component of the Oersted field:
$$\tau_{Oe}^0 = \gamma \mu_0 J_\text{e} t_{Pt}/2=\LAFO{71}{TauOe},$$
where $J_e$ is the peak current density within the Pt layer and $t_{Pt}$ is the thickness of the Pt layer.
Since the measured out-of-plane field-like torque is equal within experimental uncertainty to the torque from the Oersted field, we detect no resolvable field-like spin-orbit-torque contribution.

\section{ST-FMR sign conventions}
The extraction of artifact voltages from ST-FMR measurements depends on the sign convention used for measuring Hall voltages. A mistaken reversal of the Hall leads can be difficult to detect because, in the absence of artifacts, this does not affect the extracted torque amplitudes. However, it can lead to an incorrect artifact correction.

\begin{figure}
    \centering
    \includegraphics[width=0.5\linewidth]{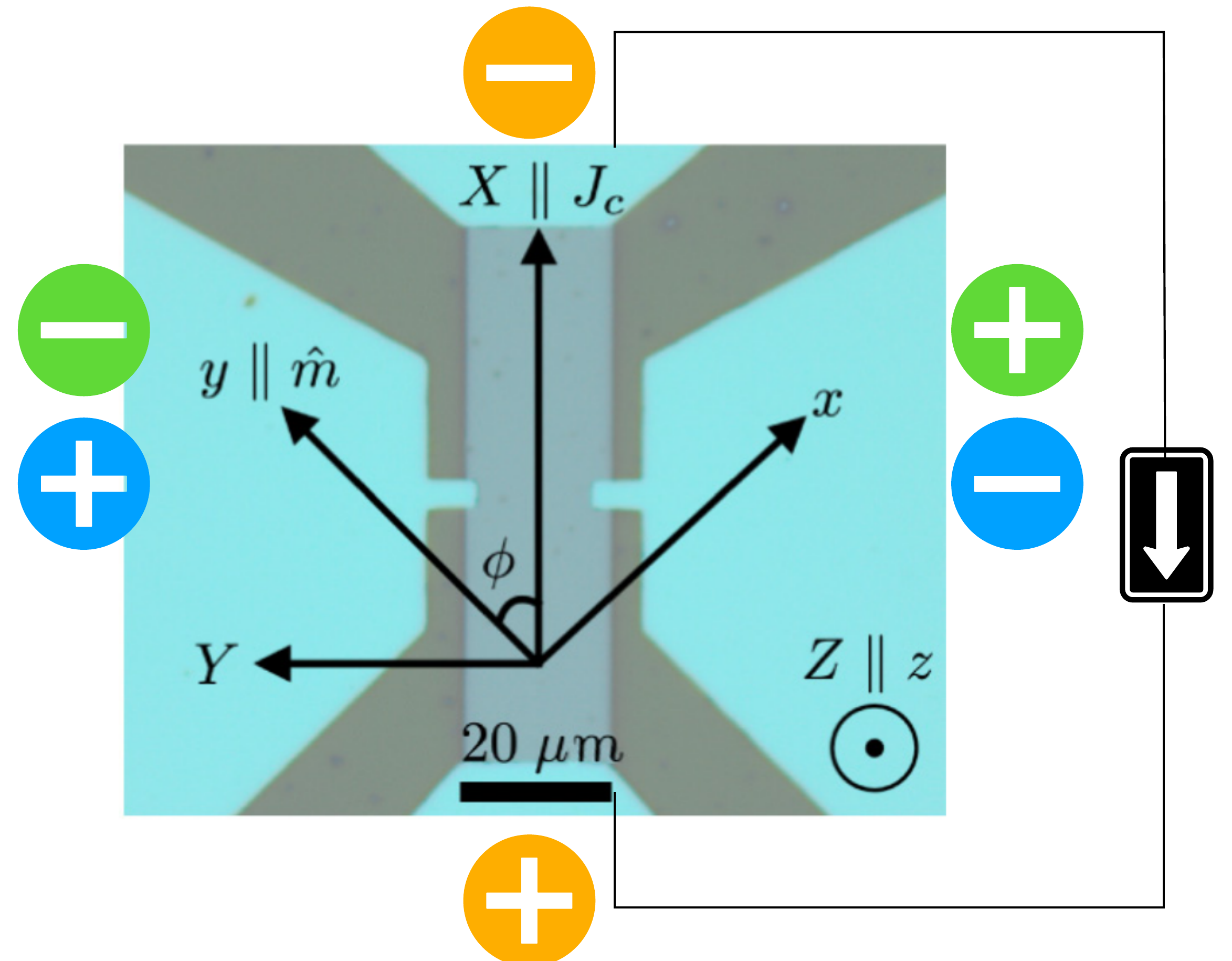}
    \caption{ST-FMR polarity conventions. Pictured is a top-down view of an ST-FMR bilayer device with magnet on top of the spin source, reproduced from reference \cite{SabaJosephHallSTFMR}. Given an applied current polarity (arrow), the colored pairs of $+$ and $-$ signs indicate different conventions for measuring voltages.  }
    \label{fig: STFMRconvention}
\end{figure}

The probe polarity conventions used throughout this manuscript are implicit in the definitions of the magnetoresistances given in the main text.
For longitudinal measurements, the expression $R_{XX}=R_0+R_{\mathrm{AMR}} m_X^2$,  along with the tacit convention that $R_0>0$, imply that the longitudinal sense leads are connected according to the orange polarity shown in Fig.~\ref{fig: STFMRconvention}.
For the Hall measurement, in this paper we have chosen the convention $R_{XY}=R_\text{PHE}m_Xm_Y+R_\text{AHE}m_Z$ with  $R_\text{PHE}$ having the same sign as $R_\text{AMR}$ corresponding to the green convention in Fig.~\ref{fig: STFMRconvention}.  The opposite convention (i.e., blue in Fig.~\ref{fig: STFMRconvention}) was used in ref.~\cite{SabaJosephHallSTFMR}. In this section, we discuss the reasons for our choice, and show how the fitting equations used for the artifact extraction depend on the convention used.

We adopt the green convention for four reasons: First, the planar Hall and the anisotropic magnetoresistance have the same sign, as mentioned above. That is, $R_{PHE} = \left(\frac{W}{L}\right)R_{AMR}$. Second, The (effective) planar Hall resistance is positive for both spin-Hall magneto-resistance and Fe, Co, Ni. Third, The anomalous Hall resistance sign matches that of the ordinary Hall effect for positive carriers. And fourth, the probe polarity with respect to the coordinate axes is consistent with the longitudinal probe.  All four signs are flipped compared with the blue convention.

Next, we discuss the impact these conventions have on the mixing voltages (Eq.~\eqref{eq: stfmr angle coef}). These equations have been copied here, but with a variable $\sigma$ to indicate which convention for the Hall voltage applies.
Under the green convention $\sigma=1$, and for the blue convention $\sigma=- 1$. It is also important that the definitions for $R_\text{PHE}$ and $R_\text{AHE}$ used in these equations also differ in sign between conventions. The longitudinal convention is fixed to the orange polarity.
$$
\begin{aligned}
S_{X X}^{\mathrm{AMR}/\mathrm{art}} & =\frac{I_{\mathrm{rf}}}{2 \alpha \omega^{+}} R_{\mathrm{AMR}} \tau_{\mathrm{DL}}^0-\frac{L}{2} E_{\mathrm{art}}^0 \\
& \equiv S_{X X}^{\mathrm{AMR}}+V_{\mathrm{art}}, \\
A_{X X}^{\mathrm{AMR}} & =\frac{I_{\mathrm{rf}}}{2 \alpha \omega^{+}} R_{\mathrm{AMR}} \frac{\omega_2}{\omega} \tau_z^0, \\
S_{X Y}^{\mathrm{PHE}/\mathrm{art}} & =-\frac{I_{\mathrm{rf}}}{2 \alpha \omega^{+}} R_{\mathrm{PHE}} \tau_{\mathrm{DL}}^0+\sigma\frac{W}{2} E_{\mathrm{art}}^0, \\
A_{X Y}^{\mathrm{PHE}} & =-\frac{I_{\mathrm{rf}}}{2 \alpha \omega^{+}} R_{\mathrm{PHE}} \frac{\omega_2}{\omega} \tau_z^0, \\
S_{X Y}^{\mathrm{AHE}/\mathrm{art}} & =\frac{I_{\mathrm{rf}}}{2 \alpha \omega^{+}} R_{\mathrm{AHE}} \tau_z^0 +\sigma\frac{W}{2} E_{\mathrm{art}}^0, \\
A_{X Y}^{\mathrm{AHE}} & =-\frac{I_{\mathrm{rf}}}{2 \alpha \omega^{+}} R_{\mathrm{AHE}} \frac{\omega_1}{\omega} \tau_{\mathrm{DL}}^0 .
\end{aligned}
$$
Notice that the only terms in these expressions that depend explicitly on $\sigma$ are the artifact terms contributing to the Hall measurements.  This is because both the measured Hall voltages (S and A coefficients) and the definitions of $R_\text{PHE}$ and $R_\text{AHE}$  switch sign upon reversing the hall probes, while the artifact electric field $E^0_\text{art}$ is defined relative to fixed coordinate axes and does not change sign depending on the convention for measuring the Hall voltage. Therefore the terms describing the contributions to the Hall voltage from the artifact electric field must contain $\sigma$ so that they properly change sign upon a change in convention for the Hall voltage.  These equations with $\sigma = -1$ correspond to the result in ref.~\cite{SabaJosephHallSTFMR} for the blue convention in Fig.~\ref{fig: STFMRconvention}.

The artifact voltage can be extracted from the following  expressions for $\eta$, where $\eta$ is defined to be proportional to the ratio of the damping-like and field-like torques (see ref.~\cite{SabaJosephHallSTFMR})
\begin{align}
\eta=\frac{-A_{X Y}^{\mathrm{AHE}}}{S_{X Y}^{\mathrm{AHE}/\text{art}}+ (-\sigma)W\left(E_{\mathrm{art}} / 2\right)}=\left\{\begin{array}{l}\frac{S_{X Y}^{\mathrm{PHE}/\text{art}}+(-\sigma)W\left(E_{\mathrm{art}} / 2\right)}{A_{X Y}^{\mathrm{PHE}}} \\ \frac{S_{X X}^{\mathrm{AMR}/\mathrm{art}}+L\left(E_{\mathrm{art}}/2\right)}{A_{X X}^{\mathrm{AMR}}}\end{array}\right..
\nonumber
\end{align}

The artifact field can therefore be determined by equating the AHE expression on the left with either of the two expressions on the right.
\begin{equation}
E_\text{art}^{top}=\frac{
\sigma\left(S_{X Y}^{\mathrm{PHE}/\mathrm{art}}+S_{X Y}^{\mathrm{AHE}/\mathrm{art}}\right) \pm \sqrt{
\left(S_{X Y}^{\mathrm{PHE}/\mathrm{art}} - S_{X Y}^{\mathrm{AHE}/\mathrm{art}}\right)^2 -4 A_{X Y}^{\mathrm{PHE}} A_{X Y}^{\mathrm{AHE}}  }
}{W} .
\nonumber
\end{equation}

\begin{equation}
    E_{\rm art}^{bottom}=\frac{\left(S_{\rm X Y}^{\mathrm{AHE}/\mathrm{art}}L-\sigma S_{\rm X X}^{\mathrm{AMR}/\mathrm{art}}W\right) \pm \sqrt{
\left(S_{\rm X Y}^{\mathrm{AHE}/\mathrm{art}}L + \sigma S_{\rm X X}^{\mathrm{AMR}/\mathrm{art}}W\right)^2 +4\sigma WL A_{\rm X Y}^{\mathrm{AHE}} A_{\rm X X}^{\mathrm{AMR}}}}{\sigma WL}.
\label{E2}
\end{equation}
Equating the PHE and AMR expressions with each other does not provide a third solution for $E_{\mathrm{art}}$, because the equivalence between the planar Hall effect and anisotropic magnetoresistance makes the resulting system singular.

Having these two ways to solve for the artifact voltage allows the degenerate solutions of these quadratic equations to be identified, while also providing a consistency check for sign-convention errors or other issues, including reversed probe polarities or lock-in phase offsets. The legitimate and degenerate solutions obtained from the two equations should each agree within experimental noise. However, the degenerate solution is more sensitive to experimental noise and may therefore differ more strongly between the two methods, allowing it to be distinguished from the legitimate solution.

\section{ST-FMR measurements with amplitude modulation of \texorpdfstring{$I_{\mathrm{rf}}$}{Irf}}

The ST-FMR expressions used here and in Ref.~\cite{SabaJosephHallSTFMR} assume an unmodulated RF drive, with $I_{\mathrm{rf}}$ defined as the peak RF current amplitude. Lock-in detection in which the RF drive is modulated allows ST-FMR measurements with improved signal-to-noise ratios. For amplitude modulation at frequency $\omega_{\mathrm{mod}}$ with 100 \% modulation depth the RF amplitude depends on time as
\[
I_{\mathrm{rf}}(t)=\bigl(1-\sin(\omega_{\mathrm{mod}} t)\bigr) I^0_{\mathrm{rf}}.
\]
At 100\% modulation depth, the RF current amplitude therefore ranges from $0$ to $2I^0_{\mathrm{rf}}$.

Because the ST-FMR mixing voltage is proportional to the RF power, it scales as $(I_{\mathrm{rf}}(t))^2$. If $V^0_{\mathrm{mix}}$ denotes the mixing voltage for an unmodulated RF current of peak amplitude $I^0_{\mathrm{rf}}$, then under 100\% amplitude modulation
\[
V_{\mathrm{mix}}(t)=V^0_{\mathrm{mix}}\bigl(1-\sin(\omega_{\mathrm{mod}} t)\bigr)^2.
\]
Expanding this expression gives
\[
V_{\mathrm{mix}}(t)
=
V^0_{\mathrm{mix}}
\left(
\frac{3}{2}
-2\sin(\omega_{\mathrm{mod}} t)
-\frac{1}{2}\cos(2\omega_{\mathrm{mod}} t)
\right).
\]
Thus the component at the lock-in reference frequency $\omega_{\mathrm{mod}}$ has peak amplitude $2V^0_{\mathrm{mix}}$, or RMS amplitude $\sqrt{2}\,V^0_{\mathrm{mix}}$. Therefore, because lock-in amplifiers generally report the RMS amplitude of the first-harmonic response, the measured lock-in signal for the ST-FMR voltage must be divided by $\sqrt{2}$ to recover the unmodulated mixing voltage denoted by $V_\text{mix}$ in the main text.

\section{Might magnon creation and annihilation affect the measurements?}
Both of the measurement approaches we have employed, lateral $+$ longitudinal ST-FMR and Sagnac interferometry, assume as part of their analyses that the magnitude of the sample magnetization is unchanged by the application of the spin-orbit torque.  However, magnon creation or annihilation can in some measurement configurations alter this magnitude \cite{noel_nonlinear_2025,noel_estimation_2025,ChenNature2024,XuAPL2025}.  We argue that magnon creation and annihilation should not interfere with obtaining quantitative results for either of the techniques we employ, for the following reasons.

The signals analyzed in ST-FMR are the resonant response to an RF current (separated from any smoothly-varying background).  The processes of magnon creation and annihilation demonstrated by No\"el et al.\ \cite{noel_nonlinear_2025,noel_estimation_2025}, in contrast, are a result of a dc current and unconnected to the resonant mode, and so they should not produce a resonance response that enters into ST-FMR analysis. We have also considered whether scattering from the uniform precessional mode might provide an alternative mechanism for exciting short-wavelength magnons.  This might produce a small decrease in the net value of the magnetization $M_s$, but a shift of $M_s$ should generate a separate artifact voltage only in the presence of a large applied dc current \cite{noel_nonlinear_2025}.  This process therefore might possibly introduce a source of error in dc-biased linewdidth analyses \cite{Ando2008,STFMRdiscrepancy_saba}, but not the forms of ST-FMR without an applied dc current.

 The creation and annihilation of magnons in response to spin-orbit torque from a quasi-static current depend on the magnetization angle as $\sin\phi$, and so these processes are most important  when the magnetization is perpendicular to the applied current \cite{noel_nonlinear_2025,noel_estimation_2025,ChenNature2024,XuAPL2025}.  Optical measurements of quasi-static magnetic deflection due to damping-like torque (including our Sagnac signal) are maximized when the magnetization is parallel to the current and go to zero as $\phi \rightarrow \pm 90^\circ$.  Therefore, these optical techniques have the advantage that they can determine the spin-torque efficiency in a configuration where the magnon creation and annihilation should go to zero.


\end{document}

%% file: sample-values.tex
\DeclareRobustCommand{\LAFO}[2]{\pgfkeysvalueof{/lafo/#1/#2}}
\DeclareRobustCommand{\Growth}[2]{\pgfkeysvalueof{/growth/#1/#2}}
\DeclareRobustCommand{\General}[1]{\pgfkeysvalueof{/general/#1}}

\pgfkeys{
  /general/.cd,
  SrcZ/.initial = {\ensuremath{50\,\Omega}}, 
}

\pgfkeys{
  /growth/gen/.cd,
  ThkRange/.initial    = {\ensuremath{7.5\text{--}14.4\,\mathrm{nm}}},
  /growth/71/.cd,
  TargetComp/.initial  = {\ensuremath{\mathrm{Li}_{0.6}\mathrm{Al}_{0.7}\mathrm{Fe}_{1.8}\mathrm{O}_{4}}},
  SubSize/.initial     = {\ensuremath{5\times5\times0.5\,\mathrm{mm^3}}},
  LiExcess/.initial    = {\ensuremath{0.1}}, 
  LaserLambda/.initial = {\ensuremath{248\,\mathrm{nm}}},
  Fluence/.initial     = {\ensuremath{2.8\,\mathrm{J\,cm^{-2}}}},
  LaserRate/.initial   = {\ensuremath{2\,\mathrm{Hz}}},
  SpotArea/.initial    = {\ensuremath{6\,\mathrm{mm^2}}},
  O2Press/.initial     = {\ensuremath{15\,\mathrm{mTorr}}}, 
  Temp/.initial        = {\ensuremath{410\,{}^\circ\mathrm{C}}}, 
  TargetDist/.initial  = {\ensuremath{3\,\mathrm{inches}}},
  CoolPress/.initial   = {\ensuremath{100\,\mathrm{Torr}}},
}

\pgfkeys{
  /lafo/052/.cd,
    Thk/.initial   = {\ensuremath{13\,\mathrm{nm}}}, 
    Alpha/.initial = {\ensuremath{2.90\times10^{-4}}}, 
  /lafo/409/.cd,
    Thk/.initial   = {\ensuremath{7.5\,\mathrm{nm}}}, 
    PtThk/.initial = {\ensuremath{2\,\mathrm{nm}}}, 
    Alpha/.initial = {\ensuremath{7.97\times10^{-3}}}, 
}

\pgfkeys{
  /lafo/71/.cd,
  Comp/.initial              = {\ensuremath{\mathrm{Li}_{0.5}\mathrm{Al}_{0.7}\mathrm{Fe}_{1.8}\mathrm{O}_{4}}}, 
  ThkNom/.initial            = {\ensuremath{14.4\,\mathrm{nm}}}, 
  Thk/.initial               = {\ensuremath{14.4(2)\,\mathrm{nm}}}, 
  PtThkNom/.initial          = {\ensuremath{2\,\mathrm{nm}}},
  PtThk/.initial             = {\ensuremath{2.0(2)\,\mathrm{nm}}},
  Stack/.initial             = {\LAFO{71}{Comp} (\LAFO{71}{ThkNom})/Pt (\LAFO{71}{PtThkNom})},
  Meff/.initial              = {\ensuremath{-0.12(2)\,\mathrm{T}}}, 
  Ms/.initial                = {\ensuremath{164(7)\,\mathrm{kA\,m^{-1}}}}, 
  Kappa/.initial             = {\ensuremath{33(1)\,\mu\mathrm{rad}}}, 
  DevL/.initial              = {\ensuremath{60\,\mu\mathrm{m}}}, 
  ChanW/.initial             = {\ensuremath{20\,\mu\mathrm{m}}}, 
  HallW/.initial             = {\ensuremath{16\,\mu\mathrm{m}}}, 
  Frf/.initial               = {\ensuremath{3.5\,\mathrm{GHz}}},
  Prf_src/.initial           = {\ensuremath{10\,\mathrm{dBm}}},
  Irf_pk_samp/.initial       = {\ensuremath{2.06(2)\,\mathrm{mA}}}, 
  AlphaST/.initial           = {\ensuremath{0.027(4)}}, 
  ResBRange/.initial         = {\ensuremath{0.22\text{--}0.26\,\mathrm{T}}}, 
  RAMR/.initial              = {\ensuremath{1.03(4)\,\Omega}}, 
  RPHE/.initial              = {\ensuremath{0.291(5)\,\Omega}}, 
  RAHE/.initial              = {\ensuremath{0.0090(5)\,\Omega}}, 
  R0/.initial                = {\ensuremath{651(7)\,\Omega}}, 
  RhoPt/.initial             = {\ensuremath{43(4)\,\mu\Omega\,\mathrm{cm}}}, 
  SamrFit/.initial           = {\ensuremath{22.3(4)\,\mu\mathrm{V}}}, 
  AamrFit/.initial           = {\ensuremath{6.5(3)\,\mu\mathrm{V}}}, 
  SpheFit/.initial           = {\ensuremath{-9.1(4)\,\mu\mathrm{V}}}, 
  ApheFit/.initial           = {\ensuremath{-1.0(3)\,\mu\mathrm{V}}}, 
  SaheFit/.initial           = {\ensuremath{7.7(3)\,\mu\mathrm{V}}}, 
  AaheFit/.initial           = {\ensuremath{-0.4(2)\,\mu\mathrm{V}}}, 
  VPheDL/.initial            = {\ensuremath{16.8(5)\,\mu\mathrm{V}}}, 
  VAmrDL/.initial            = {\ensuremath{51(1)\,\mu\mathrm{V}}}, 
  TauDLPHE/.initial          = {\ensuremath{1.0(2)\times10^8\,\mathrm{s^{-1}}}}, 
  TauDLAMR/.initial          = {\ensuremath{8(2)\times10^7\,\mathrm{s^{-1}}}}, 
  JPt/.initial               = {\ensuremath{5.1(5)\times10^{10}\,\mathrm{A\,m^{-2}}}}, 
  XiSTPHE/.initial           = {\ensuremath{0.08(2)}}, 
  XiSTAMR/.initial           = {\ensuremath{0.07(1)}}, 
  XiST/.initial              = {\ensuremath{0.07}}, 
  XiSTWrong/.initial         = {\ensuremath{0.029(6)}}, 
  ArtifactPct/.initial       = {\ensuremath{50\,\%}}, 
  SigmaSH/.initial           = {\ensuremath{1.5(3)\times10^5\,(\hbar/2e)\,\Omega^{-1}\mathrm{m^{-1}}}}, 
  Poptic/.initial              = {\ensuremath{800\,\mu\mathrm{W}}}, 
  SettleN/.initial           = {\ensuremath{10}}, 
  FEOM/.initial              = {\ensuremath{(2\pi)\,3.34\,\mathrm{MHz}}}, 
  FDrive/.initial            = {\ensuremath{(2\pi)\,2.27\,\mathrm{kHz}}}, 
  TauSag/.initial            = {\ensuremath{1\,\mathrm{s}}}, 
  Tacquire/.initial          = {\ensuremath{1\,\mathrm{s}}}, 
  SweepN/.initial            = {\ensuremath{27}}, 
  SagIpk/.initial            = {\ensuremath{7.09\,\mathrm{mA_{peak}}}}, 
  SagIrms/.initial           = {\ensuremath{5.02\,\mathrm{mA_{rms}}}}, 
  SagJ/.initial              = {\ensuremath{1.3(1)\times10^{11}\,\mathrm{A_{rms}\,cm^{-2}}}}, 
  SagBwrong/.initial         = {\ensuremath{0.12\,\mathrm{T}}}, 
  SagBsGood/.initial         = {\ensuremath{0.15\text{--}0.22\,\mathrm{T}}}, 
  XiSagFixed/.initial        = {\ensuremath{0.0009(1)}}, 
  XiSagFloat/.initial        = {\ensuremath{0.0018(2)}}, 
  MeffSagFit/.initial        = {\ensuremath{-0.07\,\mathrm{T}}}, 
  RFHeatSlope/.initial       = {\ensuremath{54(1)}}, 
  DCHeatCoeff/.initial       = {\ensuremath{2.56(2)\times10^5}}, 
  RFTransfer/.initial        = {\ensuremath{0.103(1)}}, 
  MismatchTransfer/.initial  = {\ensuremath{0.143}}, 
  MeffFlip/.initial          = {\ensuremath{-0.015\,\mathrm{T}}}, 
  MeffResRange/.initial      = {\ensuremath{-0.15\text{ to }-0.20\,\mathrm{T}}}, 
  TauFLAMR/.initial          = {\ensuremath{1.1(2)\times10^7\,\mathrm{s^{-1}}}}, 
  TauFLPHE/.initial          = {\ensuremath{6(2)\times10^6\,\mathrm{s^{-1}}}}, 
  BOe/.initial               = {\ensuremath{0.0647(6)\,\mathrm{mT}}}, 
  TauOe/.initial             = {\ensuremath{1.14(1)\times10^7\,\mathrm{s^{-1}}}}, 
}

\pgfkeys{
  /lafo/404/.cd,
  Comp/.initial       = {\ensuremath{\mathrm{Li}_{0.5}\mathrm{Al}_{0.5}\mathrm{Fe}_{2}\mathrm{O}_{4}}},
  XiST/.initial              = {\ensuremath{0.04(1)}}, 
  Kappa/.initial      = {\ensuremath{36(2)\,\mu\mathrm{rad}}},
  XiSag/.initial      = {\ensuremath{0.019(3)}},
}

\pgfkeys{
  /lafo/traditional/.cd,
  Comp/.initial     = {\ensuremath{\mathrm{Li}_{0.5}\mathrm{Al}_{1.0}\mathrm{Fe}_{1.5}\mathrm{O}_{4}}}, 
}